\documentclass[aps,prl,twocolumn,showpacs,superscriptaddress,floatfix,reprint]{revtex4-1}
\usepackage[dvips]{graphicx}
\usepackage{color}
\usepackage{latexsym}
\usepackage{amsmath}
\usepackage{amsthm}
\usepackage{amsfonts}
\usepackage{amssymb}
\usepackage{mathrsfs}
\usepackage{natbib}

\begin{document}
\title{Ground-state cooling of a carbon nanomechanical resonator by spin-polarized current}
\author{P. Stadler}
\affiliation{Fachbereich Physik, Universit{\"a}t Konstanz, D-78457 Konstanz, Germany}
\author{W. Belzig}
\affiliation{Fachbereich Physik, Universit{\"a}t Konstanz, D-78457 Konstanz, Germany}
\author{G. Rastelli}
\affiliation{Fachbereich Physik, Universit{\"a}t Konstanz, D-78457 Konstanz, Germany}
\affiliation{Zukunftskolleg, Universit{\"a}t Konstanz, D-78457 Konstanz, Germany}
%

%
\begin{abstract}
%
%
We study the nonequilibrium steady state of a mechanical resonator in the quantum regime realized by a suspended carbon nanotube quantum dot contacted by two ferromagnets.
Because of the spin-orbit interaction and/or an external magnetic field gradient, the spin on the dot couples directly to
the flexural eigenmodes. 
Accordingly, the nanomechanical motion induces inelastic spin flips of the tunneling electrons.
A spin-polarized current at finite bias voltage causes either heating or active cooling of the mechanical modes.
We show that maximal cooling is achieved at resonant transport when the energy splitting
between two dot levels of opposite spin equals the vibrational frequency. 
Even for  weak electron-resonator coupling and moderate polarizations we can achieve ground-state cooling
with a temperature of the leads, for instance, of $T=10\omega$.
%
%
%
\end{abstract}
%

%
\date{\today}
\maketitle
%

Beyond proving  useful technologically as ultrasensitive detectors of charge \cite{Li:2007} and spin \cite{Rugar:2004}, 
nanoelectromechanical systems are also interesting to address fundamental issues 
as they can enter the quantum regime at  low temperature \cite{Armour:2002,Blencowe:2004}.
For instance,  recent experiments succeeded in approaching the quantum ground state in solid objects formed 
by a huge number of atoms \cite{Rocheleau:2009,Teufel:2011,OConnell:2010}.
Particularly interesting nanoelectromechanical systems are suspended carbon nanotube quantum dots (CNTQDs) \cite{Huettel:2009,Lassagne:2009}.  
%
They emerged as an ideal system for fundamental studies in few electron quantum dots \cite{Cao:2005} 
as, for instance, demonstrated by the coherent coupling between the electron spin and its orbital magnetic moment 
(spin-orbit interaction) \cite{Kuemmeth:2008,Steele:2013,Laird:2014}.   
In addition, suspended structures also have outstanding mechanical properties as 
carbon nanoresonators can have  frequencies  in the range $f\sim$ MHz-GHz  
and yet large quantum  zero-point fluctuations ($\delta u \sim 10$pm),  
making them ideal candidates for observing quantum mechanical effects. 
In these systems, quantized vibrational modes appear 
in low temperature transport  spectroscopy \cite{Braig:2003,LeRoy:2004,Leturcq:2009,Cavaliere:2010}.

\begin{figure}[t]
\begin{flushleft}
\begin{minipage}{0.59\columnwidth}
\includegraphics[width=1.1\linewidth,clip=true]{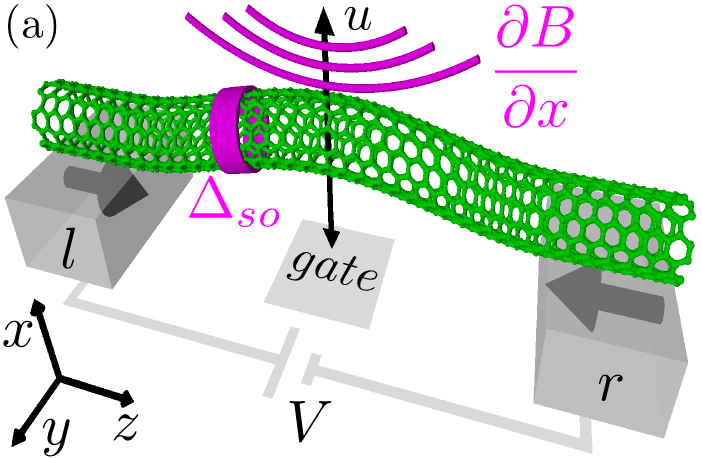}
\end{minipage}
\begin{minipage}{0.39\linewidth}
\includegraphics[width=0.85\columnwidth]{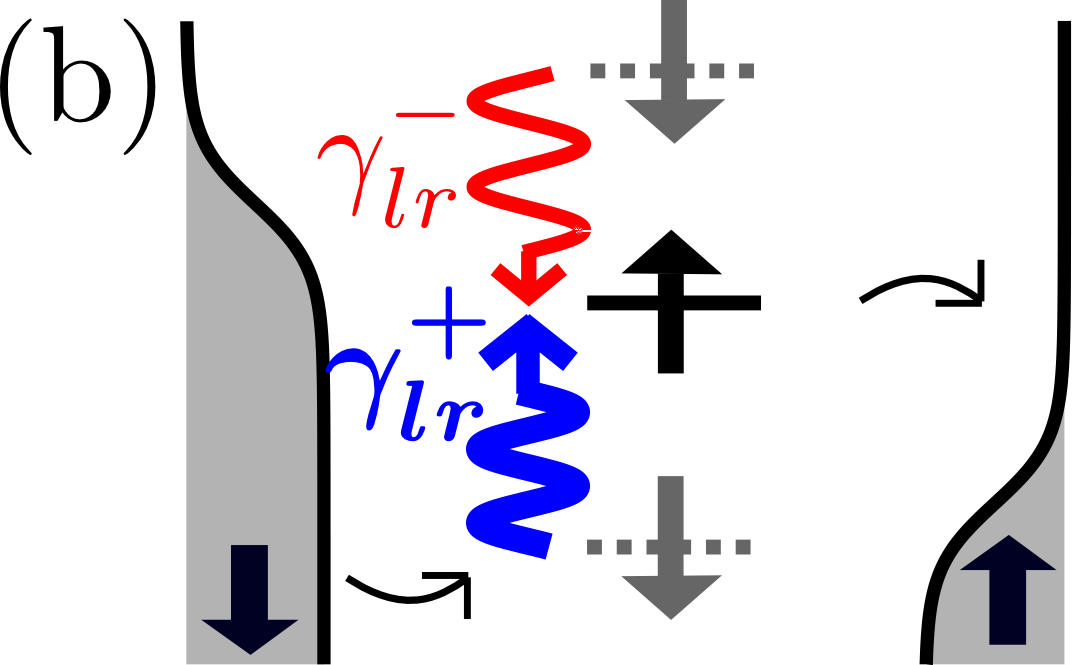}\\[3mm]
\includegraphics[width=0.85\columnwidth]{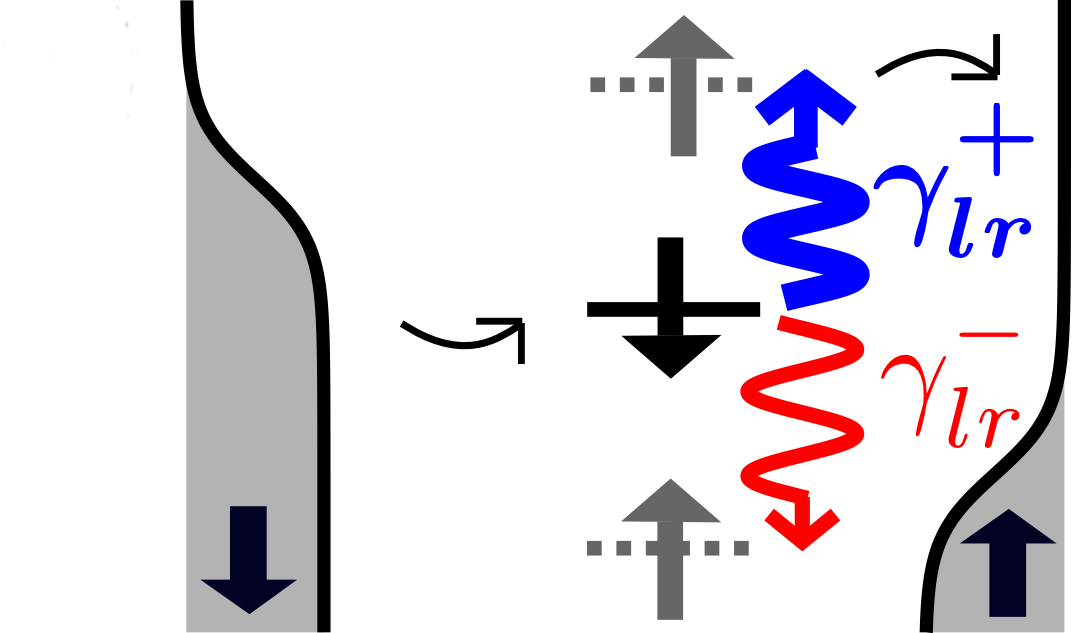}
\end{minipage}
\end{flushleft}
\caption{(color online) (a) Schematic view  of a carbon nanotube quantum dot suspended between two 
ferromagnetic leads. Because of the nanotube spin-orbit interaction and/or a magnetic field gradient, 
the dot spin's component parallel to the mechanical displacement $u$ is coupled to the flexural mode.  
(b) Examples of inelastic vibron-assisted tunneling through a single level with spin up or down (upper and lower graph).   
The spin-vibration interaction allows spin-flip tunneling through emission (red thin arrow) or absorption (blue thick arrow) of an energy quantum 
$\hbar \omega$ to or from the vibrational mode. 
These processes  are  characterized by the rates  $\gamma^{+}_{lr}$
and $\gamma^{-}_{lr}$, respectively.
}    
\label{fig:schema_experiment}
\end{figure}

Despite this amazing progress, detecting quantum signatures of flexural modes [Fig.~\ref{fig:schema_experiment}a]
still remains a challenge,  hindered by the difficulty of cooling such low-frequency 
  modes to temperatures in the quantum regime, viz., $k_BT < h f$.
Although shorter resonators with higher eigenfrequency can in principle overcome the problem \cite{Laird:2012,Island:2012}, 
cooling these modes towards their quantum ground state with phonon occupation number $\bar n \ll 1$ remains a demanding achievement. 
Even at cryogenic temperatures and with suspended nanotubes of 
length $L\sim1 \mu$m which allow flexible gate-voltage control \cite{Sahoo:2005,Benyamini:2014}, this remains a serious challenge. 
If proved feasible, such a quantum mechanical mode would be an ideal platform to test decoherence mechanisms and even exotic phenomena such as wave-function collapse theories 
in quantum states with displaced centers of mass \cite{Marshall:2003,Bassi:2013}. Another possible application is as realization of mechanical qubits in buckled carbon nanotubes \cite{Carr:2001,Werner:2004,Savelev:2006,Sillanpaa:2011}.

%
In this Letter, we show that the flexural modes can be efficiently 
cooled towards their quantum limit when a spin-polarized current is injected from 
ferromagnetic leads and when a vibrational spin-flip interaction is considered (Fig.~\ref{fig:schema_experiment}). 
Considering a flexural mode of frequency $\omega$ in a CNTQD with a quality factor   
$Q=\omega/\gamma_0 \agt 10^{4}$ ($\gamma_0$ is the mechanical damping rate) \cite{Huettel:2009,Steele:2009}, 
the resonator can be  driven towards a nonequilibrium steady state with a phonon occupation 
$\bar n =(\gamma_0 \, n_B(\omega) + \gamma \, n )/(\gamma_0 + \gamma)$,  in which 
$n_B(\omega)=1/(\exp(\omega/T)-1)$ is the thermal equilibrium occupation $(k_B=\hbar=1)$,  
and $\gamma$ and $n$ are, respectively, the damping and the effective phonon occupation induced by 
the spin-vibration interaction, which we discuss below. 

Different ways to achieve cooling of flexural modes have been analyzed \cite{Zippilli:2009,Bruggemann:2014}.  
The spin-valve that we propose has two important advantages.
First,  the spin is directly coupled to the vibration so that efficient ground-state cooling $\bar n \ll 1$ 
is achieved even for small spin polarization of the contacts. 
Second, the operating regimes, in which cooling  or heating 
of the resonator is realized,  can be controlled not only electrically but even magnetically: 
The spin valve switches from one to the other regime  either by varying the gate or the bias voltage
or either by only reversing  the magnetic polarization in one or in both ferromagnetic leads.
Such a system represents, hence, a promising candidate for the thermal control 
of nanoresonators in spintronic devices. Previous works also demonstrated that interplay between spin and nanomechanics can lead to interesting effects such as mechanical self-excitations \cite{Radic:2011ie}.

{\sl Spin-vibration interaction.}-
The system is sketched in Fig.~\ref{fig:schema_experiment}(a).
For a single flexural mode $n$  with frequency $\omega_n$ and oscillating along the $x$ axis, 
suspended CNTQDs are characterized by a spin-vibration interaction of the form   
$\hat{H}_{n} = \lambda_n^{\phantom{g}} \hat{\sigma}_x^{\phantom{g}}  (\hat{b}_n^{\phantom{g}}+\hat{b}^{\dagger}_n)$  
in which $\hat{\sigma}_x$ is the component of the spin operator (Pauli matrix) parallel to the mechanical motion
and $\hat{b}_n$ $(\hat{b}^{\dagger}_n)$  is the bosonic creation (annihilation) operator  associated with the harmonic mode.
This kind of interaction can be achieved extrinsically or intrinsically. 

In the first case, the interaction arises from the relative motion of the suspended nanotube in a magnetic gradient  
added to the homogeneous magnetic field $B$ in a similar setup as used, e.g., in 
magnetic resonance force microscopy experiments \cite{Rugar:2004,Bargatin:2003,Rabl:2009} or  
in magnetized microcantilevers coupled to nitrogen vacancy centers in diamond \cite{Arcizet:2011,Kolkowitz:2012}.  
For  small harmonic oscillations,  one obtains $\lambda_n \simeq \mu_B  (\partial B_x/\partial x) X_n$ 
with $\mu_B$ the Bohr magneton, $\partial B_x/\partial x$ the average gradient along the tube's axis, 
$X_n=u_n  \langle f_n(z) \rangle $ the  amplitude of the single vibrational mode with $u_n=1/\sqrt{2m\omega_n}$,  
$f_n(z)$ the mode waveform,  and  $\langle\cdots\rangle$ the average over the electronic orbital in the dot.
We estimated  $\lambda=0.5$ MHz for the fundamental (even) mode  with  
$\partial B_x/\partial x  = 5 \cdot 10^{6}$ T/m 
\cite{Xue:2011,Supplemental_Material}\nocite{Flensberg:2010,Landau1986,Poggio:2010}. 

In the second case, the spin-orbit coupling  due to 
the circumferential orbital motion mediates the interaction between 
the electron spin and the flexural modes \cite{Rudner:2010,Palyi:2012,Ohm:2012}. 
In the one-orbital (valley) subspace,  the interaction coupling constant  reads 
$\lambda_n \simeq (\Delta_{SO}/2)  dX_n/dz$ with 
$\Delta_{SO}$ the spin-orbit coupling constant and 
$dX_n/dz=u_n  \langle df_n(z)/dz \rangle $ \cite{Supplemental_Material}. 
In this case, one can estimate $\lambda \sim 2.5$MHz for the first odd mode \cite{Palyi:2012,Supplemental_Material}.
We notice that for a quantum dot formed in a nanotube with symmetric orbital electronic density, 
the two  interactions discussed here couple vibrational modes  of different parity. 
Other microscopic mechanisms lead also to similar coupling \cite{Borysenko:2008}.

In the presence of magnetic fields, the four-level structure  of a single quantum dot shell can be tuned.
In particular, close  to a crossing point, it is possible to have two levels of 
opposite spin and the same orbital so that their energy separation is smaller than the temperature $T$ or the bias voltage $V$  and yet larger than the energy distance from other levels  \cite{Steele:2013,Palyi:2012,Supplemental_Material} .
Focusing on the transport on this two-level subspace, we consider the  model Hamiltonian
\begin{equation}
\label{eq:Hamiltonian_int}
\hat{H}
= 
\hat{H}_{l}+\hat H_d
+
\lambda (d^\dagger_+ d_-+d^\dagger_- d_+)
 ( \hat{b}^{\dagger} + \hat{b}^{\phantom{}} )
+
\omega \hat{b}^{\dagger}  \hat{b}^{\phantom{}}   \, ,
\end{equation}
in which the dots part reads $\hat H_d= \sum_{\sigma} \varepsilon_{\sigma}^{\phantom{\dagger}}  
\hat{d}^{\dagger}_{\sigma} \hat{d}_{\sigma}^{\phantom{}}$ and the lead part reads 
\begin{equation}
\label{eq:Hamiltonian_el}
\hat{H}_{l} 
= 
\sum_{\alpha\sigma k} 
\left[
\varepsilon_{k\sigma}^{\phantom{\dagger}}
\hat{c}^{\dagger}_{\alpha\sigma k}
\hat{c}_{\alpha\sigma k}^{\phantom{\dagger}}
+
\left( 
t_{\alpha\sigma}^{\phantom{\dagger}}
\hat{c}^{\dagger}_{\alpha\sigma k} 
\hat{d}_{\sigma}^{\phantom{\dagger}}  + 
\textrm{H.c.} 
\right) 
\right].
\end{equation}
The operators $\hat{c}^{\dagger}_{\alpha\sigma k}$ and 
$\hat{d}^{\dagger}_\sigma$   are creation operators for the 
electronic states $k$ in the $\alpha=l,r$ (left, right)  leads and the dot states with spin $\sigma=\pm$.  
The latter have energy $\varepsilon_{\sigma} = \varepsilon_0 + \sigma \varepsilon_z/2$ 
with the energy separation $\varepsilon_z$.
The ferromagnets are magnetized in the $z$ directions and their effect on the spin-polarized tunneling is captured in spin-dependent tunneling rates $\Gamma_{\alpha}^{\sigma} = \pi |t_{\alpha\sigma}|^2 \rho_{\alpha\sigma}$. 
Here, $\rho_{\alpha\sigma}$ denotes the spin-$\sigma$ density of states at the Fermi level of lead $\alpha$, and $t_{\alpha\sigma}$ the tunneling amplitude, and we can define a polarization $p_{\alpha}=(\Gamma_{\alpha}^{+} - \Gamma_{\alpha}^{-})/(\Gamma_{\alpha}^{+} + \Gamma_{\alpha}^{-})$. 

{\sl  Results.}-
In the regime of weak spin-vibrational interaction,  
electrons tunneling from the leads to the dot yield a (small) renormalization of the vibration frequency  
and a damping of the mechanical motion with friction coefficient $\gamma$. 
In addition, at finite bias voltage, the electron current  drives the mechanical oscillator to a steady nonequilibrium regime with a phonon occupation 
$n$. 
To determine these quantities, we employ the Keldysh Green functions technique 
to calculate the phonon propagator $D(t,t')=-i \langle T_{\mathcal{C}}\hat{u}(t)\hat{u}(t') \rangle$,   
where $T_{\mathcal{C}}$ denotes the time-ordering operator on the Keldysh contour $\mathcal{C}$  \cite{Rammer:2007}. 
We have solved the Dyson equation with the self-energy associated with the spin-vibration interaction Eq.~(\ref{eq:Hamiltonian_el}) to the leading order in $\lambda$. 
This approximation is sufficient for $\gamma\ll \omega $  \cite{Mitra:2004}.
We find 
$ \gamma=\sum_{\alpha\beta s} s \, \gamma^{s}_{\alpha \beta} $ ($s=\pm 1$)
and for the occupation
\begin{equation}
\label{eq:main_results}
	n= \frac{1}{\gamma}
	\sum_{\alpha\beta s} s \gamma^{s}_{\alpha\beta} 
	n_B(\omega+s(\mu_{\alpha}-\mu_{\beta}))\,.
\end{equation}
Here, we introduced the lead chemical potentials $\mu_{\alpha}$  and
\begin{equation}
\label{eq:rates}
\gamma^{s}_{\alpha\beta}  = 
\frac{\lambda^2}{2}   \int \frac{d\varepsilon}{2\pi} 
T_{\alpha\beta}^s(\varepsilon,\omega)
 f_{\alpha}(\varepsilon)\left[1\mathord- f_{\beta}( \varepsilon+ s\omega)\right] \, ,
\end{equation}
with  the Fermi function  
$f_{\alpha}(\varepsilon) \mathord= {\left\{ 1 + \exp\left[ (\varepsilon-\mu_{\alpha}) /T \right] \right\}}^{-1}$,
$T_{\alpha\beta}^s(\varepsilon,\omega)=
\sum_{\sigma}
L^{\sigma}_{\alpha}(\varepsilon)L^{-\sigma}_{\beta}(\varepsilon + s\omega)$ and
  $L_\alpha^{\sigma}(\varepsilon)$ $=$ $ 2\Gamma_\alpha^\sigma/$ $\left[(\Gamma_{l}^\sigma+\Gamma_{r}^{\sigma})^2+(\varepsilon-\varepsilon_\sigma)^2\right]$.

The essential point of our proposal is that $z$ (or $y$) spin polarized electrons injected in the dot 
are perpendicular to the spin component coupled to the nanotube oscillations ($x$ axis) so that spin-flip transitions are  needed to exchange energy with the vibrational mode. 
These inelastic processes are characterized by the rates $\gamma^{s}_{\alpha\beta}$ Eq.~(\ref{eq:rates})  describing
a spin flip of an electron tunneling from lead $\alpha$ to lead $\beta$ accompanied by the absorption $(s=+)$ or emission $(s=-)$ of an energy quantum of the vibron.
The weighted sum gives the total damping coefficient $\gamma$. 

On the one side, for the parallel configuration of the ferromagnets $p_lp_r > 0$, we always found heating of the oscillator at finite bias voltage and we will not further consider this case.
On the other side, for the antiparallel configuration, we obtain heating and efficient cooling  also for different polarizations $|p_l| \neq |p_r|$. We found similar results even in the limit of one unpolarized lead (see discussion below).
Hereafter, we restrict our discussion to the antiparallel configuration with the same polarization 
$p_r= p$ and $p_l=-p$  with sgn($p$)=sgn($\varepsilon_z$).
We note that the inverted polarizations with sgn($p$)=-sgn($\varepsilon_z$) 
is equivalent to a reversed voltage.
Depending on the sign of the voltage, we also found a strong overheating 
of the mechanical resonator $\bar n  \gg 1$ for which the system approaches
 an instability region with a negative damping $\gamma < 0$. 
This configuration corresponds to the operating regime in which phonon lasing has been discussed recently \cite{Khaetskii:2013}. 
Electromechanical instability was also obtained in a different microscopic model based on the magnetomotive interaction between current and vibration in Ref. \cite{Radic:2011ie} in which it was shown that the feedback action of the vibration on the current can lead to mechanical self-excitations in a suspended CNT-QD contacted to a single ferromagnet.
In the remainder of the Letter we consider antiparallel magnetizations with $p>0$, $\varepsilon_z>0$, and $V>0$. 

For  $T  \gg  \Gamma^{\sigma}_{\alpha}$, one can use an analytic approximation for the rates $\gamma^{s}_{\alpha\beta}$, which is in excellent agreement with the full results  Eq.~(\ref{eq:rates}).
The analysis of such an incoherent regime can also be addressed by using a Pauli master equation \cite{Pistolesi:2009ks}.
The Lorentzian functions  appearing in Eq.~(\ref{eq:rates})  
can be treated separately as $\delta$ functions in the integral and we can cast each rate as the sum of two rates 
$\gamma^{s}_{\alpha\beta} \simeq \sum_{\sigma} \gamma^{s\sigma}_{\alpha\beta}$,  
for tunneling through the dot level $\sigma$, respectively.
They read
\begin{eqnarray}
\label{eq:approximation}
\gamma_{\alpha\beta}^{s\sigma} &=& \frac{\lambda^2}{\Gamma_l^{\sigma}+\Gamma_r^{\sigma}}
\left\{ 
\Gamma_{\alpha}^{\sigma}\Gamma_{\beta}^{-\sigma}  T_{+}^{s\sigma}
 f_{\alpha}(\varepsilon_{\sigma})\left[1- f_{\beta}(\varepsilon_{\sigma}+s\omega)\right] 
\right. \nonumber \\
& +&  \left. 
\Gamma_{\alpha}^{-\sigma} \Gamma_{\beta}^{\sigma}  T_{-}^{s\sigma}
 f_{\alpha}(\varepsilon_{\sigma}-s\omega)\left[1- f_{\beta}(\varepsilon_{\sigma})\right]  
\right\}
\end{eqnarray}
with $T_{\pm}^{s\sigma} = 1/\left[(\Gamma_l^{-\sigma}+\Gamma_r^{-\sigma})^2+(\sigma\varepsilon_z\pm s\omega)^2\right]$.

\begin{figure}[tbp]
\includegraphics[width=\linewidth,angle=0.]{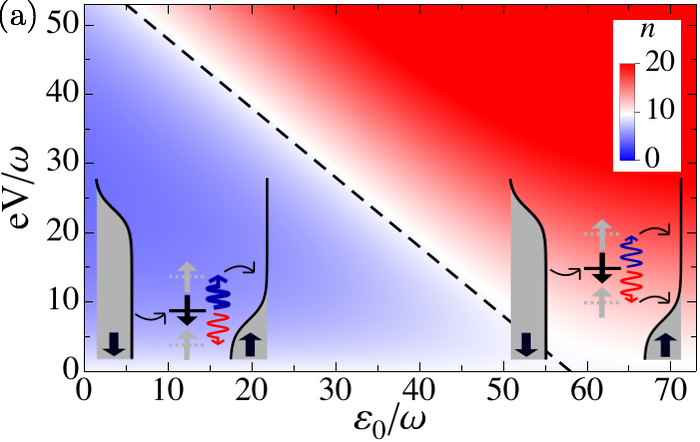} \hfill\\[-0mm]
\includegraphics[width=\linewidth,angle=0.]{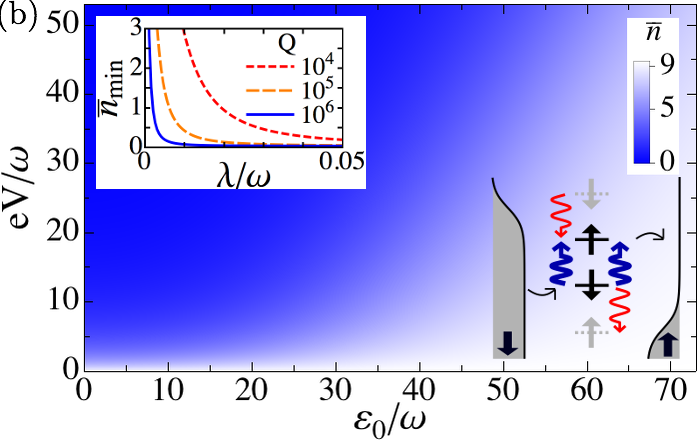} \hfill\\[-4mm]
\caption{(color online) 
Phonon occupation as function of  the bias voltage $V$ and $\varepsilon_0$. 
The parameters are $p=1$ (fully polarized), $\Gamma^{-}_l = \Gamma^{+}_r = 0.2 \omega$,
and $T= 10 \omega $. 
White corresponds to $n_B(\omega)$.  
(a) Vanishing external damping $\gamma_0=0$, $\varepsilon_z = 10 T$,
$\mu_r = \varepsilon_0 - eV$, and $\mu_l=\varepsilon_0$.
The black dashed line indicates the transition from cooling to heating (see text). 
Inset:  schematic behavior  of the relevant spin-flip processes in the region of cooling (left) and heating (right).  
(b) Resonant regime $\varepsilon_z = \omega$ with
$\gamma_0=10^{-5}\omega$, $\lambda/\omega=0.01$, and $\mu_{l,r}=\varepsilon_0 \pm eV/2$. 
Inset, left: the minimum occupation $\bar n_{\textrm{min}}$
as a function of the spin-vibration coupling constant $\lambda$ for different quality factors.
Inset, right: schematic behavior of the energy levels and of the inelastic resonant slip-flip tunneling.
}    
\label{fig:fully_polarized}
\end{figure}
{\sl  Fully polarized contacts.-}
To gain  insight into the problem,  we describe in detail the  case of 
fully polarized ferromagnets although 
efficient cooling is achieved even for small polarizations.
For $p=1$,  the diagonal rates vanish $\gamma^{s}_{ll}=\gamma^{s}_{rr}=0$,
as the electron cannot come back to its original lead after a spin flip.  
Moreover,  in the high-voltage limit $e V \gg T$ $(e>0)$,  we can safely neglect the processes $\gamma_{rl}^{s} \simeq 0$ being $V>0$ 
as electrons tunneling from the right lead are Pauli blocked. 
Accordingly, the total damping reduces to the sum of only two processes $\gamma \simeq \gamma^{+}_{lr} - \gamma^{-}_{lr}$ and 
the expression of $n$ simplifies to the average distribution resulting from these two competing processes
\begin{equation}
\label{eq:highV}
n \simeq \frac{\gamma^{+}_{lr} n_B(\omega+eV) - \gamma^{-}_{lr} n_B (\omega-eV)}{\gamma^{+}_{lr}-\gamma^{-}_{lr}}   
\simeq \frac{\gamma^{-}_{lr}}{ \gamma^{+}_{lr}-\gamma^{-}_{lr}} \, .
\end{equation}
The second step in Eq.~($\ref{eq:highV}$) holds for  $eV \gg  \omega$, when 
the nonequilibrium phonon occupation is completely ruled by the ratio $\gamma^{+}_{lr}/\gamma^{-}_{lr}$.
Although in the region of stability defined by $\gamma^{+}_{lr} >\gamma^{-}_{lr}$ the total damping is always positive, $n$ can show heating or cooling:  for $\gamma^{+}_{lr} \agt \gamma^{-}_{lr} $  the mechanical oscillator is almost undamped 
and it is actively heated to $n \agt n_B(\omega)$ whereas for  $\gamma^{+}_{lr}\gg \gamma^{-}_{lr}$ 
the dominant emission processes yield an efficient cooling of the oscillator, viz. $ n\ll n_B(\omega)$.
This is the main mechanism of cooling underlying our proposal.

We now discuss the result for the fully polarized case in Fig.~\ref{fig:fully_polarized}.
Since $\Gamma^{+}_{l}=\Gamma^{-}_{r}=0$ one of two terms appearing in Eq.~(\ref{eq:approximation}) 
vanishes for each spin channel.
For symmetric contacts $\Gamma_l^{-}=\Gamma_r^{+}=\Gamma$ and setting 
$T^s = \lambda^2 \Gamma/[ \Gamma^2 + {\left(\omega -s \varepsilon_z \right)}^2]$, the single spin-channel 
rates read
\begin{equation}
\label{eq:gamma_delta}
\gamma^{s\sigma}_{lr} 
= 
T^{s}
\, 
f_l (\varepsilon_{\sigma} - s \omega \delta_{\sigma+}) 
\left[ 1 -  f_r(\varepsilon_{\sigma}+  s \omega \delta_{\sigma-})  \right] \, .
\end{equation}
In Fig.~\ref{fig:fully_polarized}(a)  we show an example of the case $\varepsilon_z \gg \omega$ 
for an asymmetric voltage bias for which only the spin-down level is involved in transport. 
In this limit  $T^{+} \simeq  T^{-} $  
and  the difference between the  absorption  rate $\gamma^{+}_{lr}$ and emission rate $\gamma^{-}_{lr}$ is mainly given by the 
product of the electronic occupations in Eq.~(\ref{eq:gamma_delta}).
The system is expected to switch from cooling to heating when we move from the regime $\gamma_{lr}^+\gg\gamma_{lr}^-$ to the regime $\gamma_{lr}^+\agt\gamma_{lr}^-$. In a simple picture, the switch is expected close to the line $\mu_r=\varepsilon_{-}$. For $\mu_r>\varepsilon_{-}$ (cooling region) the emission processes are suppressed due to the occupation of the low energy level in the right lead [left inset Fig.\ref{fig:fully_polarized}(a)]. For $\mu_r < \varepsilon_{-}$ (heating region) emission processes are relevant and they compete with the absorption ones  [right inset Fig.\ref{fig:fully_polarized}(a)]. At finite temperature, the thermal broadening of the Fermi functions causes a smooth transition between the two regimes so that the crossing line corresponding to $n=n_B$ occurs at $\varepsilon_-=\mu_r+8T^2/\varepsilon_z$ to leading order in $T/\varepsilon_z$ and for $T \gg \omega$. Note that, in this discussion, the left lead plays only the role of a source for injecting one electron with spin up in the dot level. Hence cooling is achieved even for a normal left contact ($p_l=0$).

The minimum of the phonon occupation as a function of voltage decreases with 
the ratio $\varepsilon_z/\omega$.
The optimal cooling is achieved at
$\varepsilon_z=\omega$. At this point and in the limit $eV \gg (T,\omega,\varepsilon_0)$, 
$f_l \simeq 1$ and $f_r \simeq 0$ and the phonon occupation of Eq. \eqref{eq:main_results} 
becomes $n \simeq {( \Gamma/\omega)}^2$.
Further decreasing the ratio $\varepsilon_z/\omega$ does not improve the cooling.

The strong cooling obtained for the resonant regime can be explained as follows. 
The absorption processes for each spin channel are now the same and we have  
$\gamma^{++}_{lr}=\gamma^{+-}_{lr}$ 
as the virtual levels $\varepsilon_{-}+\omega$ and $\varepsilon_{+}-\omega$,  
which are involved in the spin-flip tunneling for cooling,  coincide, respectively, 
to the real dot spin levels $\varepsilon_{+}$ and  $\varepsilon_{-}$. 
This yields  a strong enhancement of the (transmission) function $T^{+}$ (phonon absorption)
as compared to $T^{-}$  (phonon emission)  in Eq.~(\ref{eq:gamma_delta}), namely $T^+\gg T^{-}$,
which explains the strong cooling effect. 
As a consequence, $n$ has a weak dependence on the alignment of the average level position $\varepsilon_0$ 
and the lead chemical potential $\mu_{\alpha}$. 
In Fig.\ref{fig:fully_polarized}(b) we show  the resonant case with a finite intrinsic damping 
$\gamma_0/\omega=10^{-5}$ to illustrate the behavior of $\bar{n}$.

\begin{figure}[tbp]
\includegraphics[width=0.9\linewidth,angle=0.]{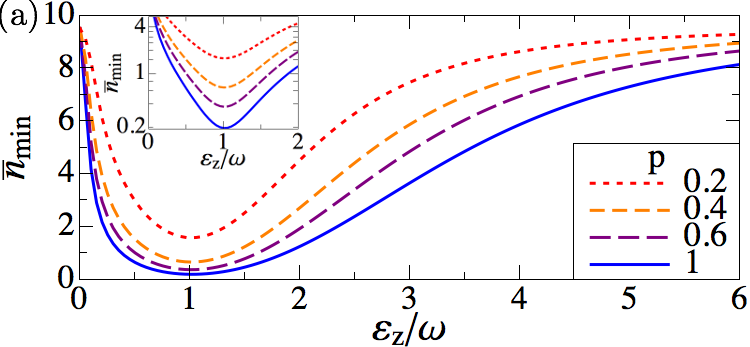} 
\includegraphics[width=0.9\linewidth,angle=0.]{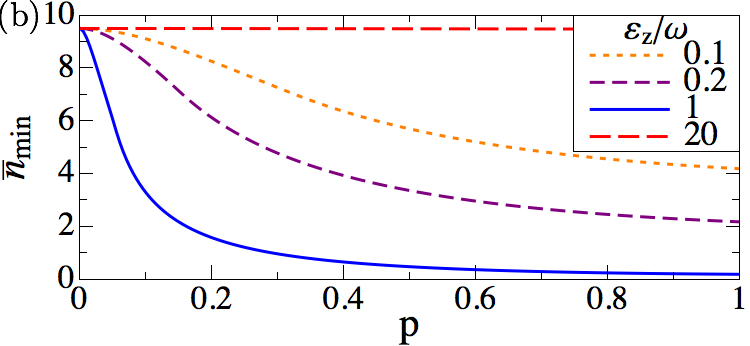}
\caption{(color online). 
Minimum of the phonon occupation $\bar{n}_{min}$ for an 
 intrinsic damping of $Q=10^4$ and $\lambda/\omega=0.05$. The temperature 
 is $T=10\omega$ and $\Gamma_l=\Gamma_r=0.2\omega$. (a) Minimal occupation as function of 
 $\varepsilon_z/\omega$ for different polarizations. 
 The inset (log scale for $y$ axis) shows that maximal cooling is achieved at $\varepsilon_z=\omega$. 
 (b) Minimal phonon occupation as a function of polarization for
 different energy separations.  
}
\label{fig:occupation}
\end{figure}

{\sl Effect of finite polarization.-}
We  discuss now the effect of finite polarization. 
For symmetrically applied voltage,
the results for the minimal value $\bar{n}_{min}$ as a function of the energy
separation for different polarizations are shown in Fig.~\ref{fig:occupation}(a).
Even in this case, at arbitrary fixed polarization, optimal cooling is again achieved for the resonant regime $\omega=\varepsilon_z$. 
A finite polarization always reduces the minimum occupation as $\bar{n}_{min}$ decreases as a 
function of $p$ independent of the ratio $\varepsilon_z/\omega$ [Fig. \ref{fig:occupation}(b)]. 
To discuss this behavior we consider the analytic high-voltage approximation  for the phonon occupation given by 
\begin{equation}
\label{eq:n_finite_pol}
n \simeq  
\frac{
\gamma^{-}_{lr} + n_B(\omega) \left( \gamma_{ll}^{\phantom{g}} + \gamma_{rr}^{\phantom{g}}  \right)
} 
{
\gamma^{+}_{lr}  -  \gamma^{-}_{lr} + \gamma_{ll}^{\phantom{g}} +\gamma_{rr}^{\phantom{ll}} 
},
\end{equation}
where we set the short notation $\gamma_{\alpha\alpha}=\gamma_{\alpha\alpha}^{+}-\gamma_{\alpha\alpha}^{-}$.
From Eq.~(\ref{eq:n_finite_pol}) we observe that the diagonal lead processes $\gamma_{\alpha\alpha}$, which are not present for $p=1$, 
have the effect of thermalizing the oscillator.  
As an example, assuming a strong asymmetry of the leads,  as for instance $\Gamma_{l} \simeq 0$ ($\Gamma_{r} \simeq 0$),
we have $\gamma^{\pm}_{lr}=0$: the dot is contacted only with one left (right) lead and the oscillator is always at the thermal equilibrium.
Such processes compete with the cooling processes $\gamma_{lr}^{+}$ leading to an increase of the minimum phonon occupation.

%

Clearly, taking into account the intrinsic damping of the mechanical oscillator also increases the minimum phonon occupation.
Remarkably,  a phonon occupation of $\bar{n}_{min}\simeq 0.5$ is still achieved for $Q \simeq 10^{4}$, $\lambda/\omega=0.05$, 
 and polarizations $p>0.48$ [Fig.\ref{fig:occupation}(b)]. 
The minimal phonon occupation reduces to $\bar{n}_{min}=0.2$ at $p=1$. 
An occupation of $\bar{n}_{min}\simeq 0.5$ is also obtained for $Q \simeq 10^{5}$ and $p>0.3$
($\bar{n}_{min}=0.05$ at $p=1$).
Motivated by a recent experiment that reported  large spin-orbit interaction coupling $\Delta_{SO}$   \cite{Steele:2013}, 
one can also consider coupling constants of order  $\lambda/\omega=0.2$ which implies a strong reduction of the polarization required 
for cooling. As an example, $\bar{n}_{min} \simeq 0.5 $ for   $Q \simeq 10^{4}$ and $p>0.3$.
Therefore, we conclude that even for modest polarizations, which appears feasible in promising experiments with CNT-QDs  
\cite{Sahoo:2005,Cottet:2006,Jensen:2005},  quantum ground-state cooling  is achievable.

{\sl  Conclusions.-}
 In summary, we discussed a suspended CNTQD  forming a nanomechanical spin valve with a direct coupling 
 between the dot spin and the flexural modes  showing that ground-state cooling is achievable 
 with moderated spin-current polarization.

\acknowledgments
We thank O. Arcizet, V. Bouchiat,  G.\ Burkard, A.\ K.\  H{\"u}ttel, E.\ Scheer, E.\ Weig, and W. Wernsdorfer for useful and stimulating discussions.
This research was kindly supported by the EU FP7 Marie Curie Zukunftskolleg Incoming Fellowship Programme, University of Konstanz (Grant No. 291784) and the DFG through SFB 767 and BE 3803/5.

\bibliography{references}

\onecolumngrid
\section*{Supplemental material for: Ground state cooling of a carbon nano-mechanical resonator  by  spin-polarized current}

\section{Microscopic model}
We describe the microscopic model for the spin-vibration interaction 
in a suspended carbon-nanotube quantum dot  mediated by the spin-orbit coupling or by a magnetic gradient.
Concerning the electronic model and the interaction mediated spin-orbit coupling, the derivation of the 
effective Hamiltonian with the spin-deflection coupling  has been already reported in literature 
and we address to Refs.\cite{Rudner:2010,Flensberg:2010,Palyi:2012}  for more  details. 
In addition we  extend  the analysis to  take into account the effect of a magnetic gradient.
A possible way to produce a magnetic gradient and oppositely polarized leads is to use a proper-shaped nano-magnet polarized in x direction placed perpendicular above the center of nanotube at the right distance. In this way the nano-magnet could polarize the two leads oppositely along z-direction (i.e. the same orientation of the magnetic field lines in these spaces) and meanwhile, the nanomagnet produces a magnetic field gradient along x direction.

\subsection{Spin Hamiltonian for a carbon nanotube quantum dot}
In quantum dot carbon nanotubes, the transversal as well as the  longitudinal 
 orbital motion are quantized  due to a confining potential along the carbon nanotube axis $z$. 
Such a spectrum is thus discrete with four-fold degeneracy associated to the 
real spin and the orbital (valley).
For such a dot-shell, we use the  notation $\left| \tau, \sigma \right>$ with $\tau =\pm$ (isospin) 
and $\sigma=\pm$ (real spin) for the unperturbed eigenstates with  the spin quantization  along the $z$-axis. 
The spin-orbit interaction and the magnetic field remove this degeneracy.
Projecting these interactions onto the states $\left| \tau, \sigma \right>$ of the  Hilbert subspace,  
the effective low-energy Hamiltonian reads \cite{Rudner:2010,Flensberg:2010,Palyi:2012}
\begin{equation}
\label{eq:H_eff_dot}
H_d =  
\frac{\Delta_{SO}}{2} \tau_3 {\bf t}(z) \cdot   \boldsymbol{ \sigma} 
- \mu_{orb} \tau_3  {\bf B} \cdot  {\bf t}(z) 
+  \mu_B    {\bf B}  \cdot \boldsymbol{ \sigma} 
+
\Delta_{KK'} \tau_1 \, ,
\end{equation}
which is valid owing to the energy scale separation between 
the high-energy spacings associated to the longitudinal quantization and the circumferential
quantization (energy gap) and the coupling energies appearing in Eq.~(\ref{eq:H_eff_dot}).
$\mu_{orb}$ and $\mu_B$ are respectively the magnetic  orbital momentum and Bohr magneton,  
${\bf B}$ is the external magnetic field, $ \boldsymbol{ \sigma}=(\sigma_x,\sigma_y,\sigma_z)$ and  $(\tau_1,\tau_2,\tau_3)$ are 
the Pauli matrices associated to the real spin and to the isospin. 
$\Delta_{SO}$ and $\Delta_{KK'}$ are the spin-orbit and inter valley coupling matrix elements.
${\bf t}(z)$ is the local tangent unit vector at each point of the tube. 
Moreover the  inter-valley coupling  is generally a small effect $\Delta_{KK'} \ll (\Delta_{SO}, \mu_{orb}B,\mu_B B)$, 
and we can neglect it hereafter as long as our discussion does not involve the crossing point of 
the levels with different orbital  quantum number.

\subsection{Mechanical oscillator}
For a freely suspended elastic rod of small radius  compared to its length $(R\ll L)$, 
the Hamiltonian for the mechanical flexural motion in one plane reads \cite{Landau1986}
\begin{equation}
\label{eq:H_mec}
H_{vib}= \int^{L}_{0}\!\!\!\!\! dz \left[ \frac{\mbox{   }\, \hat{\pi}^2(z)}{2\rho_L}
+
\frac{E I}{2} {\left( \frac{d^2\hat{u}(z)}{dz^2} \right)}^2 
-
\frac{T}{2}   {\left( \frac{d\hat{u}(z)}{dz} \right)}^2
\right]
=
\sum_n \hbar \omega_n \hat{b}_n^{\dagger} \hat{b}_n^{\phantom{}} \, ,
\end{equation}
with $\hat{u}(z)$ the operator corresponding to the local transversal displacement, $\hat{\pi}(z)$ 
the conjugate operator $[\hat{u}(z),\hat{\pi}(z)] = i\hbar \delta(z-z')$, $E$ the carbon nanotube Young's modulus,
$I=\pi R^4/4$ the transversal inertial moment and 
$T$ is the uniform tension on the rod.
The second form of Hamiltonian Eq.~(\ref{eq:H_mec}) is obtained via a canonical transformation 
in which  the local displacement reads  $\hat{u}(z) =\sum_n f_n(z) u_n ( \hat{b}_n^{\phantom{g}}  + \hat{b}_n^{\dagger} )$ with $f_n$ 
the profile function and $u_n=\sqrt{\hbar/(2m\omega_n)}$ and $m$ the nanotube's mass, 
and $\hat{\pi}(z) =i \sum_n  f_n(z) m\omega_n u_n (\hat{b}_n^{\dagger} -  \hat{b}_n^{\phantom{g}})$.
For sufficiently strong tension $T \gg (\pi/4) E R^4/L^2$, one can neglect the second term in 
Eq.~(\ref{eq:H_mec}) corresponding to the elastic energy for the length-variation of an infinitesimal element 
caused by the local bending.
Then  the wave-form functions $\left\{ f_n(z) \right\}$ are the orthonormal set solutions of the standard 
wave-equation for an elastic string with eigenfrequencies $\omega_n=(n+1)\,\pi \sqrt{T/(\rho_L L^2)}$,
and $f_n(z)= \sqrt{2} \sin[ \pi (n+1)  z/L ]$ for integers $n \geq 0$.

\subsection{Spin-vibration interaction}
A direct coupling  arises from local changes in the direction of the nanotube axis  ${\bf t}(z)$ in the 
laboratory reference frame which operates even for zero external magnetic field \cite{Rudner:2010,Flensberg:2010,Palyi:2012}.
Here we discuss this mechanism, generalized to the presence of a magnetic gradient.
In this case, a local deflection causes two effects:  
i)  a local variation of the nanotube axis $\delta{\bf t}(z)  \simeq (d\hat{u}(z)/dz,0,0)$  \cite{Rudner:2010,Flensberg:2010} and 
ii) a variation of the magnetic field seen by the single electron $\delta{\bf{B}}=(\partial{\bf{B}}/\partial x) \hat{u}(z)$.
For small amplitudes, we approximate the product  ${\bf B} \cdot {\bf t}(z) \simeq B_z
+ {\bf B} \cdot \delta{\bf t}(z)  + \delta {\bf B}   \cdot {\bf t}(z) $ 
in which we can neglect $\delta{\bf t}(z) \cdot \delta {\bf B} $ corresponding  to higher-order terms in $\hat{u}$.
Hence we cast the Hamiltonian as the sum of four terms:
\begin{equation}
\label{eq:H_0_dot}
H_d = \tau_3 \left( \frac{\Delta_{SO}}{2}  s_z   - \mu_{orb}  B_z \right)
+  
\mu_B \left(   {\bf B}  \cdot \boldsymbol{ \sigma}  \right) 
+  H_{\sigma,\tau} + H_{\sigma}  + H_{\tau}  \, ,
\end{equation}
with the first term of Eq. \eqref{eq:H_0_dot} is the unperturbed Hamiltonian. 
Projecting on a single flexural eigenmode $n$ in the $x-$axis, 
the interacting terms  read 
\begin{eqnarray}
H_{\sigma,\tau} &=&  
 \frac{\Delta_{SO}}{2}  \left< f'_n \right>   u_n \left( \hat{b}_n^{\phantom{g}}  + \hat{b}_n^{\dagger} 
\right)   \tau_3  \sigma_x \, ,
\\
H_{\sigma}  &=&\mu_B \sum_{\nu=x,y,z}  \left< \frac{\partial B_{\nu}}{\partial x} f_n \right> 
 u_n \left( \hat{b}_n^{\phantom{g}}  + \hat{b}_n^{\dagger} \right) \sigma_{\nu} 
 \simeq
 \mu_B   \frac{\partial B_x}{\partial x}  \left<  f_n \right> 
 u_n \left( \hat{b}_n^{\phantom{g}}  + \hat{b}_n^{\dagger} \right) \sigma_{x} \, ,
  \\
H_{\tau}  &=&
\!\! -\mu_{orb}    \left< B_x f'_n  +  \frac{\partial B_z}{\partial x}  f_n \right>   u_n   \left( \hat{b}_n^{\phantom{g}}  + \hat{b}_n^{\dagger} \right) \tau_3
\equiv 0
\qquad \left(\mbox{for} \quad B_x=\partial B_z/\partial x=0\right)   \, ,
\end{eqnarray}
with $f'_n=df_n/dz$  and $\left< \dots \right> = \left< \phi(z)(\right|  \dots \left| \phi(z) \right>  $ 
the  average  against the  longitudinal  orbital wave function. 
$ H_{\sigma,\tau} $ is  the intrinsic spin-vibration interaction mediated by the spin-orbit coupling,  
whereas $ H_{\sigma} $ corresponds to the  coupling  induced by the magnetic gradient.
To be define, we consider only one component and we set $dB_y/dx=dB_z/dx=0$. 
The last term $H_{\tau}$ is the interaction of the vibration with the orbital part which vanishes for 
$B_x=0$ and $dB_z/dx=0$.
Moreover,  for a magnetic gradient with its leading components $dB_x/dx$  perpendicular to the 
nanotube $z-$axis , we can neglect the variation of the magnetic field along this axis
$d B_{\nu}/dz \simeq 0$  so that we extracted the gradient from 
the  orbital average. 

We observe that, for symmetrically distributed charge of the single electronic orbital,    
the interaction $H_{\sigma,\tau}$ and $H_{\sigma}$ are alternatively determined only by parity of the modes. 
Using a simple  flat density ${|\phi(z)|}^2 = 1/(\epsilon L)$ extending on a length scale $\epsilon L$,  
we estimate 
\begin{equation}
\left< f_n\right> 
\simeq
\frac{2\sqrt{2}}{\pi (n+1)} \left(  \frac{\sin\left[ (n+1) \epsilon \pi/2\right]}{\epsilon}  \right) 
\sin\left[ (n+1 )\pi/2 \right]
\, ,
\qquad
\left< f'_n \right> 
\simeq
\frac{2\sqrt{2}}{L} \left(  \frac{\sin\left[ (n+1) \epsilon \pi/2\right]}{\epsilon}  \right) 
\cos\left[ (n+1) \pi/2 \right] 
\, .
\end{equation}
Taking $\epsilon \simeq 1$, we have 
$\left< f_0 (z)\right>  =2\sqrt{2}/\pi$ for the  first  even mode (fundamental) 
and $\left< df_1 (z)/dz\right>  =2\sqrt{2}/L$ for the first odd mode.
Assuming  $\Delta_{SO} \simeq 400 \mu$eV, $u_0 =10$pm and the length $L=0.5 \mu$m, 
we obtain the coupling constant in $\hat{H}_{\sigma,\tau}$ as 
$\lambda=(\Delta_{SO}/2) \left< df/dz \right> u_0 \simeq 194 \cdot 10^{-4} \mu$eV corresponding to a frequency of $\lambda/h  = 2.5 MHz$.
For mechanical oscillator of frequency $f =100 MHz$, 
the dimensionless damping constant corresponds to  $\lambda/(h f)=\lambda/(\hbar \omega)=0.025$.
On the other hand, for $\hat{H}_{\sigma}$,  
we  have  $\lambda=  \mu_B (dB/dx) \left< f \right> u_0 \simeq 30 \cdot 10^{-4} \mu$eV
for a gradient $dB_x/dx=5 \cdot 10^{6} T/m$ \cite{Poggio:2010,Xue:2011},  
and it  corresponds to a frequency of $\lambda/h  = 0.45 MHz$.

\subsection{Spin level doublet}
As follows we discuss two specific examples which realize the model Hamiltonian $H_{el}$ in the main text.
Specifically, we discuss  a spectrum with pairs of spin eigenstates (doublets) corresponding to the two subspaces 
$\tau=\pm$ in which $H_{\sigma,\tau}$ and $H_{\sigma}$ are diagonal. 
These spin doublets are  parallel to the ferromagnets polarization  (longitudinal along $z$ or transversal along $y$ axis)    
allowing spin-polarized current to flow through the dot.
\paragraph{Longitudinal polarization.}
We consider an uniform  magnetic field $B_z$ applied along the nanotube axis $z$. 
We have longitudinal polarization of the ferromagnetic and  the  eigenstates of the unperturbed dot Hamiltonian Eq.~(\ref{eq:H_0_dot}) are 
exactly the basis $\left|\tau,\sigma \right>$.
Each orbital subspace $\tau=\pm$ has two pair of  spin-eigenstates parallel to the ferromagnets.  
Far away from the crossing points, the level degeneracy between two opposite spin-states 
and different orbitals is completely removed and one can focus on the transport through a single level. 
Approaching the magnetic field around $B^*_z \simeq \Delta_{SO}/(2\mu_B)$  
a spin doublet is almost degenerate \cite{Rudner:2010,Palyi:2012}. 
Tuning the magnetic field around this point, we achieve the situation discussed in the main text for which two levels are involved 
in the transport and strong cooling is obtained at resonance $\varepsilon_z=\omega$.

\paragraph{Quasi-transversal polarization.}
For a uniform magnetic field with leading transversal  component $B_y \gg B_z$, 
the ferromagnet are transversally polarized and the dot levels are also, with good approximation, eigenstates of the spin 
along the same direction $y$  $(B_y \gg \Delta_{SO}/2\mu_B)$.  
For  these nanotubes with negligible spin-orbit $\Delta_{SO} \simeq \mu_B B_z$ ($B_z\sim 2$mT)
 the spin-vibration coupling  arises from a magnetic gradient. 
Moreover, owing to finite $B_z$, the degeneracy is still removed.
For instance, for  $B_y \sim 15$mT and $B_z \sim 2$mT, 
we have a orbital subspace $(\tau=-)$ with two opposite spin level tunable 
around the frequency $\varepsilon_z \sim \omega \sim 100$MHz and separation from the other spin-doubles of 
$\mu_{orb} B_z \sim$ GHz.

\section{Dyson equation and Phonon Self energy for the spin-vibration coupling}

We use the Keldysh-Green function technique to calculate the phonon occupation calculating 
the self-energy of the phonon propagator to the  order  $\lambda^2$ which corresponds to the leading 
order in the spin-vibration coupling  \cite{Rammer:2007,Mitra:2004}. 
We refer to the notation in Ref.~\onlinecite{Rammer:2007}.
The retarded and Keldysh phonon propagators are defined as $(t >0)$
\begin{equation}
D^R(t)  = -\frac{i}{\hbar} \theta(t) \left<  \left[ \hat{u}(t),   \hat{u}(0)\right] \right>  \, , \qquad 
D^K(t)  = -\frac{i}{\hbar} \theta(t) \left<  \left\{\hat{u}(0), \hat{u}(t)\right\} \right>  \, ,
\end{equation}
in which $[\, , \, ]$ ($\{\, , \, \}$) denotes the  commutator (anti-commutator).
In the frequency space, using the triangular Larkin-Ovchinnikov representation, the triangular matrix $\check{D}(\omega)$ satisfies the following Dyson equations 
\begin{equation}
 \label{KeldyshPhononProp}
\check{D}(\varepsilon) =  \check{d}(\varepsilon)  +  \check{d}(\varepsilon) \check{\Pi}(\varepsilon) \check{D}(\varepsilon) \, , 
\end{equation}
in which  the free bare phonon propagator  reads $d^{R,A}(\varepsilon)=2\hbar \omega/\left((\varepsilon\pm i\eta)^2+(\hbar\omega)^2\right)$ and 
$d^{K}(\varepsilon)=-2\pi i (\delta(\varepsilon-\hbar\omega)+\delta(\varepsilon+\hbar\omega))\mathrm{coth}(\hbar \omega/(2k_BT)))$ with an infinitesimal small real part $\eta$.
To the first leading order $\lambda^2$ in the spin-vibration interaction,  
${\Pi}^{R,A}_{\lambda^2}$ and ${\Pi}^K_{\lambda^2}$ read 
\begin{eqnarray}
\Pi^R_{\lambda^2}(\varepsilon) &=& -i\frac{\lambda^2}{2}\int\frac{d\varepsilon^\prime}{2\pi}\sum_\sigma \left[ G_{-\sigma}^{K}(\varepsilon^\prime) 
G_{\sigma}^A(\varepsilon^\prime-\varepsilon)  + G_{-\sigma}^R(\varepsilon^\prime) G_{\sigma}^{K}(\varepsilon^\prime-\varepsilon) \right] \, , \label{eq:pi_R} \\
\Pi^K_{\lambda^2}(\varepsilon) &=&
-i\frac{\lambda^2}{2}\int\frac{d\varepsilon^\prime}{2\pi}\sum_{\sigma} \left[
G_{-{\sigma}}^K(\varepsilon^\prime) G_{\sigma}^K(\varepsilon^\prime-\varepsilon)  
+ G_{-{\sigma}}^R(\varepsilon^\prime)
G_{\sigma}^A(\varepsilon^\prime-\varepsilon)+ G_{-{\sigma}}^A(\varepsilon^\prime) G_{\sigma}^R(\varepsilon^\prime-\varepsilon)
\right]  \, . \label{eq:pi_K}
\end{eqnarray}
Note that the interaction vertex due to the spin-vibration couples only spins of opposite sign. The electron Keldysh Green's function of the dot appearing Eqs.~(\ref{eq:pi_R}),(\ref{eq:pi_K}) are the Green functions of the unperturbed Hamiltonian Eqs.$\sim$(1,2) 
of the paper corresponding to exact solvable problem of a dot-level coupled to the leads.
In the wide band approximation, they read
$G^{R,A}_{\sigma}(\varepsilon)=1/[\varepsilon-\varepsilon_\sigma\pm i (\Gamma_{l}^{\sigma}+\Gamma_{r}^{\sigma})]$
and   \begin{math} G_\sigma^K = 2iG^R_\sigma \left(\Gamma_l^\sigma (2 f_l-1)+\Gamma_R^\sigma (2 f_r-1) \right)G^A_\sigma \end{math}. 
Solving Eq.~\eqref{KeldyshPhononProp}, one obtains 
\begin{eqnarray}
D^{R}(\varepsilon) 
&=& 
\frac{2\hbar\omega}{\varepsilon^2-(\hbar\omega)^2-2\hbar\omega\Pi^{R}_{\lambda^2}(\varepsilon)}
\simeq
\frac{1}{\varepsilon-\hbar(\omega+\Delta\omega)+i\gamma}-\frac{1}{\varepsilon+\hbar(\omega+\Delta\omega)+i\gamma} \, , \\
D^K(\varepsilon) &=&D^R(\varepsilon) \Pi^K(\varepsilon) D^A(\varepsilon) \simeq \frac{\pi}{\gamma} \, \Pi^K_{\lambda^2}(\varepsilon) \left[\delta(\varepsilon-\hbar\omega)+\delta(\varepsilon+\hbar\omega)\right].
\end{eqnarray}
As the interaction is small, we expanded the retarded phonon propagators around $\varepsilon\lesssim \hbar \omega$ 
and the damping is given by $\gamma= - \mbox{Im} [\Pi^R(\omega)]$ whereas the frequency renormalisation is
 $\Delta \omega= \mbox{Re} [\Pi^R(\omega)]$.
The phonon occupation is obtained from $n= (i/8\pi) \int d\varepsilon D^K(\varepsilon)- 1/2$.

\section{Phonon self-energy for the vibration-environment coupling}

We consider a mechanical oscillator coupled to the environment which is described as 
an ensemble of harmonic oscillator (the Caldeira-Leggett model)
\begin{equation}
\hat{H} = \hbar\omega \hat{b}^{\dagger} \hat{b} +   (\hat{b}^{\dagger} +  \hat{b}) 
\sum_n \lambda_n (\hat{b}^{\dagger}_n +  \hat{b}_n)
+
\sum_n \hbar \omega_n  \hat{b}^{\dagger}_n \hat{b}_n \,.
\end{equation}
As the Hamiltonian is quadratic, the model is exactly solvable: The phonon self-energy is composed by only one irreducible diagram. 
For instance, in the frequency space, the retarded self energy is given by
\begin{equation}
\Pi_{en}^R(\varepsilon) = \sum_n \lambda_n^2 \left(\frac{1}{\varepsilon-\hbar\omega_n+i\eta}-\frac{1}{\varepsilon+\hbar\omega_n+i\eta}\right) \, .
\end{equation}
To mimic the dissipation, the ensembles of oscillators form a bath with a continuos spectrum.
Then, by replacing the sum with an integral over the frequencies and  approximating $\Pi_{en}^R(\varepsilon)\simeq \Pi_{en}^R(\omega) $, we obtain 
\begin{eqnarray}
\gamma_0  &=& -\mbox{Im }\Pi_{en}^R(\omega)  = \omega/Q \\      
\Pi_{en}^K(\omega) &=& 2i\mbox{Im }\Pi_{en}^R(\omega)\mathrm{coth}(\omega)  \, ,
\end{eqnarray}
with  $Q$ the quality factor of the oscillator. 
The phonon occupation is evaluated by inserting the two contributions, the phonon self energies 
of the environment  $\Pi_{en}$ and spin-vibration coupling $\Pi_{\lambda^2}$ into the Dyson equation  Eq.~(\ref{KeldyshPhononProp}).
Finally, one obtains
\begin{equation}
\bar{n}   =\frac{\gamma_0 \, n_B + \gamma \, n }{\gamma_0 + \gamma}.
\end{equation}
with the Bose distribution function $n_B=n_B(\omega)$.

\section{Derivation of the rates expressions}
The formula for the phonon occupation $n$ of the paper is obtained by the following steps. 
First, the second and third term of Eq. \eqref{eq:pi_K} vanish. Second, in the first term of Eq. \eqref{eq:pi_K},
we rewrite the products of the Fermi functions
$(2f_\alpha(\varepsilon^\prime+\varepsilon)-1)(2f_\beta(\varepsilon^\prime)-1)) $ as
$2(f_\alpha(\varepsilon^\prime+\varepsilon)-f_\beta(\varepsilon^\prime)) \mathrm{coth}(\varepsilon-\mu_\alpha+\mu_\beta)+1$.
As  last step, we use a shift in the integration and we get the formula
\begin{equation}
\textrm{Im }\Pi_{\lambda^2}^K(\varepsilon) = -\lambda^2 \int \frac{d\varepsilon^\prime}{2\pi} \sum_{\alpha\beta} \sum_s sT_{\alpha\beta}^s(\varepsilon^\prime) f_\alpha(\varepsilon^\prime) \left(1+2n_B\left(\frac{\varepsilon-\mu_\alpha+\mu_\beta}{k_BT}\right)\right)
\end{equation}
with the transmissions 
$T_{\alpha\beta}^s(\varepsilon) = \sum_\sigma L_\alpha^\sigma(\varepsilon) L_\beta^{-{\sigma}}(\varepsilon+s\omega)$
and
$L_\alpha^{\sigma}(\varepsilon) \mathord= 2\Gamma_\alpha^\sigma/\left[(\Gamma_{l}^\sigma+\Gamma_{r}^{\sigma})^2+(\varepsilon-\varepsilon_\sigma)^2\right]$.

The formula for the damping coefficient $\gamma=-\textrm{Im} \Pi^R_{\lambda^2}$ is calculated as follows. 
In Eq. \eqref{eq:pi_R} we multiply the first term with 
$G^R_\sigma(\varepsilon^\prime-\varepsilon) {G^R_\sigma(\varepsilon^\prime-\varepsilon)}^{-1}$ 
and the second term with
${G^A_{-\sigma}}^{-1}(\varepsilon^\prime)G^A_{-\sigma}(\varepsilon^\prime) $.
The imaginary part of the retarded polarization can then be written as
\begin{equation}
\textrm{Im }\Pi^R_{\lambda^2}(\varepsilon) =  \frac{\lambda^2}{2} \int \frac{d\varepsilon^\prime}{2\pi} \sum_{\alpha\beta} \sum_s sT_{\alpha\beta}^s(\varepsilon^\prime) f_\alpha(\varepsilon^\prime).
\end{equation} 
The expression of the main paper is then obtained by adding and subtracting 
$T_{\alpha\beta}^sf_\alpha(\varepsilon)f_\beta(\varepsilon+\omega)$ 
and by using the relation
$T_{\alpha\beta}^{s}(\varepsilon-\omega)=T_{\beta\alpha}^{-s}(\varepsilon)$.



\end{document}